\documentclass{icrc2009}

\usepackage{graphicx}   
\usepackage[caption=false]{caption}    
\usepackage[font=footnotesize]{subfig} 
\usepackage{fixltx2e}
\usepackage{url}

\newcommand{\shorttitle}[1]%
{\markboth{Proceedings of the 31\MakeLowercase{$^{st}$} ICRC, {\L}\'{o}d\'{z} 2009}{#1} }
\newcommand{\etal}{\MakeLowercase{\textit{et al. }}} 

\begin{document}
\title{Study of High $p_{T}$ Muons in IceCube}

\author{\IEEEauthorblockN{Lisa Gerhardt\IEEEauthorrefmark{1}\IEEEauthorrefmark{2} and
			  Spencer Klein\IEEEauthorrefmark{1}\IEEEauthorrefmark{2}
                           for the IceCube Collaboration\IEEEauthorrefmark{4}}
                            \\
\IEEEauthorblockA{\IEEEauthorrefmark{1}Lawrence Berkeley National Laboratory, Berkeley, California 94720}
\IEEEauthorblockA{\IEEEauthorrefmark{2}University of California, Berkeley, Berkeley, California 94720}
\IEEEauthorblockA{\IEEEauthorrefmark{4}For a full list see http://www.icecube.wisc.edu/collaboration/authorlists/2009/4.html}}

\shorttitle{Gerhardt \etal High $p_{T}$ Muons in IceCube}
\maketitle

\begin{abstract}
Muons with a high transverse momentum ($p_{T}$) are produced in cosmic ray air showers via semileptonic decay of heavy quarks and the decay of high $p_{T}$ kaons and pions. These high $p_{T}$ muons have a large lateral separation from the shower core muon bundle. IceCube is well suited for the detection of high $p_{T}$ muons. The surface shower array can determine the energy, core location and direction of the cosmic ray air shower while the in-ice array can reconstruct the energy and direction of the high $p_{T}$ muon. This makes it possible to measure the decoherence function (lateral separation spectrum) at distances greater than 150~meters. The muon $p_{T}$ can be determined from the muon energy (measured by dE/dx) and the lateral separation. The high $p_{T}$ muon spectrum may also be calculated in a perturbative QCD framework; this spectrum is sensitive to the cosmic-ray composition.

  \end{abstract}

 
\section{Introduction}
The composition of cosmic rays with energies above 10$^{6}$ GeV is not well known. At these energies, the flux of cosmic rays is so low that direct detection with balloon or satellite-borne experiments is no longer feasible and indirect measurement with larger ground arrays must be used. These arrays measure the electronic and hadronic components of a cosmic ray air shower, and must rely on phenomemological interaction models to relate observables like muon and electron density to primary composition. These interaction models are based on measurements made at accelerators that reach a maximum energy roughly equivalent to a 10$^{6}$ GeV proton cosmic ray \cite{engel}. Extrapolation to the energy of the detected cosmic rays leads to uncertainties in the composition of cosmic rays at high energies. An alternate method of determining the composition is to use muons with a high transverse momentum ($p_{T}$) \cite{klein}.
 
At transverse momenta on the order of a few GeV/c, the muon $p_T$ spectrum can be calculated using perturbative QCD (pQCD).  Such calculations have been made for RHIC and the Tevatron, and the data is in quite good agreement with modern fixed order plus next-to-leading log calculations \cite{Ramona}.  These experimental studies give us some confidence in pQCD calculations for air showers.  

Most of the high-energy muons that are visible in IceCube are produced from  collisions where a high-energy (large Bjorken$-x$) parton interacts with a low Bjorken$-x$ parton in a nitrogen or oxygen target.  These collisions will produce heavy (charmed and bottom) quarks and also jets from high $p_T$ partons; the jets will fragment into pions and kaons.  Pions and kaons produce ``conventional'' muons that have a soft spectrum roughly proportional to E$^{-3.7}$ \cite{gaisser} and typicallly have a low $p_{T}$. In contrast, charm and bottom quarks are produced early in the shower. The resulting muons (``prompt'' muons) have a harder spectrum, a higher $p_{T}$, and are the dominant source of atmospheric muons above 10$^5$ GeV \cite{pasq}. 

If these particles are produced in the forward direction (where they will be seen by ground based detectors), they can carry a significant fraction of the energy of the incident nucleon.  The muon energy and $p_T$ can be related to the energy of the partons that make up the incident cosmic ray. For example, a $10^{16}$ eV proton has a maximum parton energy of $10^{16}$ eV, while the maximum parton energy for a $10^{16}$ eV iron nucleus (with A=56) is $1.8\times10^{14}$ eV. These two cases have very different kinematic limits for high $p_T$, high-energy muon production, so measurements of high $p_T$ muon spectra are sensitive to the cosmic-ray composition.  

High $p_T$ muons constitute a small fraction of the muons visible in IceCube; the typical muon $p_T$ is of order a few hundred MeV/c.  In this domain of ``soft physics,'' the coupling constants are very large, and pQCD calculations are no longer reliable. One is forced to rely on phenomenological models; different hadronic interaction models give rather different results on composition \cite{creview}.  

IceCube, a kilometer-scale neutrino telescope, is well suited for the detection of muons with $p_{T}$s above a few GeV/c \cite{klein}. When completed in 2011, it will consist of a 1~km$^3$ array of optical sensors (digital optical modules, or DOMs) buried deep in the ice of the South Pole and a 1~km$^2$ surface air shower array called IceTop. IceTop has an energy threshold of 300~TeV and can reconstruct the direction of showers with energies above 1~PeV within $\sim$1.5$^\circ$ and locate the shower core with an accuracy of $\sim$10~m \cite{klepser}. The in-ice DOMs (here referred to as IceCube) are buried in the ice 1450~m under IceTop in kilometer-long strings of 60 DOMs with an intra-DOM spacing of 17~m. IceCube can reconstruct high quality tracks of high energy muons with sub-degree accuracy. IceTop can measure the energy, position, and direction of the shower and IceCube can do the same for the muons. These values can be used to calculate the $p_{T}$:
   \begin{equation}
     p_{T}=\frac{d E_{\mu}}{hc}
    \label{pt_eqn}
   \end{equation}
where $E_{\mu}$ is the energy of the high $p_{T}$ muon, $d$ is its lateral separation and $h$ is the interaction height of the shower, here taken as an average value of 25 km. The interaction height loosely depends on the composition and a full treatment of this is planned in the future. Taking 150~m, 25 meters more than the separation between strings of DOMs in IceCube, as a rough threshold for the two-track resolution distance of the high $p_{T}$ muon from the shower core gives a minimum resolvable $p_{T}$ of 6 GeV/c for a 1~TeV muon. 

Multiple scattering and magnetic fields can deflect the muons as they travel to the IceCube detector, but this is only equivalent to a few hundred MeV worth of $p_{T}$ and is not a strong effect in this analysis. The high $p_{T}$ muon events are near-vertical which gives them a short slant depth. In order for muons to reach the IceCube detector they must have an energy of about 500~GeV at the surface of the Earth. These higher energy muons are deflected less by multiple scattering and the Earth's magnetic field.

The combined acceptance for cosmic ray air showers which pass through both IceTop and IceCube is 0.3~km$^2$sr for the full 80-string IceCube array \cite{bai}. By the end of the austral summer of 2006/2007, 22 IceCube strings and 26 IceTop tanks had been deployed. The combined acceptance for showers that pass through both IceTop-26 and IceCube-22 is 0.09 km$^2$sr. While this acceptance is too small to expect enough events to generate a $p_{T}$ spectrum, it offers an excellent opportunity to test reconstruction and background rejection techniques.

\section{Previous Measurement of High $p_{T}$ Muons}
MACRO, an underground muon detector, has previously measured the separation between muons in air showers with energies ranging roughly from 10$^4$ GeV to 10$^6$ GeV \cite{ambrosio}. MACRO measured muon separations out to a distance of about 65 meters. The average separation between muons was on the order of 10 m, with 90\% of the muons found with a separation of less than 20~m \cite{ambrosio}. MACRO simulated air showers and studied the muon pair separations as a function of the $p_{T}$ of their parent mesons. They verified the linear relationship between $p_{T}$ and separation shown in Eq. \ref{pt_eqn} (with a small offset due to multiple scattering of the muons) out to momenta up to 1.2 GeV/c.

\section{High $p_{T}$ Rate Estimations}
The decay of charm will produce roughly 10$^6$ prompt muons/year with energies in excess of 1~TeV inside the 0.3 km$^2$sr combined acceptance of the full IceCube array \cite{thunman}.  
Different calculations agree well in overall rate in IceCube, but, at high (1~PeV) energies, different parton distribution functions lead to rates that differ by a factor of 3 \cite{enberg}. 

We can roughly estimate the fraction of muons that are produced at high $p_T$ using PYTHIA $pp$ simulations conducted for the ALICE muon spectrometer at a center of mass energy of 14~TeV, which indicate that roughly 0.6\% of these events will have a $p_{T}$ of at least 6 GeV/c \cite{valle}.  These simulations are done at a higher center of mass energy than the bulk of the IceCube events, and also were for muons produced at mid-rapidity (and, therefore, higher projectile $x$ values and lower target $x$ values than in IceCube).  However, they should be adequate for rough estimates.  These estimated rates of high $p_{T}$ muons are further reduced by approximately a factor of 10 by requiring that the high $p_{T}$ muon be produced in coincidence with a shower that triggers IceTop, leaving a rough expectation of $\sim$500 events per year with a $p_{T}$ greater than 6 GeV/c in the 80-string configuration of IceCube. The estimated number of events above a given $p_{T}$ from the decay of charm is given in Table \ref{table_pt} neglecting the uncertainties mentioned above. At higher $p_T$, bottom production may also contribute significantly to muon production. 
  \begin{table}[]
  \caption{Estimated number of events from charm above different $p_{T}$ thresholds in 1 year with the 80-string IceCube array}
  \label{table_pt}
  \centering
  \begin{tabular}{|c|c|c|}
  \hline
   $p_{T}$ [GeV/c] &  Separation [m] & Number \\
   \hline 
    $\ge$6 & 150 & $\sim$500 \\
    $\ge$8 & 200 & $\sim$200 \\
    $\ge$16 & 400 & $\sim$5 \\
  \hline
  \end{tabular}
  \end{table}

The rate of 1~TeV muons from conventional flux is expected to be about 10$^9$ events/year in the full IceCube configuration. The vast majority of these muons will have a low $p_{T}$. Based on the $p_{T}$ spectrum measured from pions produced in 200 GeV center of mass energy $pp$ collisions by the PHENIX collaboration, the number of events expected with $p_{T} >$ 6 GeV/c is estimated to be 1 in 6 $\times$ 10$^6$ \cite{adler}. Requiring a $p_{T} >$ 6 GeV/c and an IceTop trigger leaves roughly 20 events/year in the full IceCube configuration from the conventional flux of muons.

\section{Reconstruction Methods}
High $p_{T}$ muons will appear as a separate track coincident in time and parallel with the track from the central core of low $p_{T}$ muons. Generally the bundle of low $p_{T}$ muons leaves more light in the detector than the high $p_{T}$ muon. 

\begin{figure}[]
  \centering
  \includegraphics[width=3.5in]{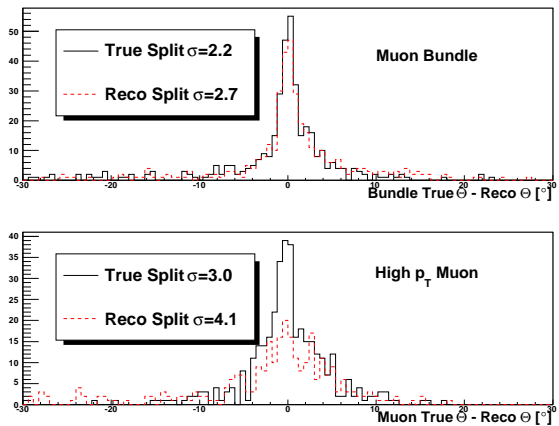}
  \caption{Zenith angle resolution of the high $p_{T}$ reconstruction algorithm. The sigma are the results of Gaussian fits to the distributions.}
  \label{fig_reco}
 \end{figure}

Current reconstruction algorithms in IceCube are designed to reconstruct single tracks. In order to reconstruct these double-track events the activated DOMs (i.e. DOMs that detect at least one photon) are split into groups using a k-means clustering algorithm \cite{mackay}. The first group is the muon bundle and the second group is the high $p_{T}$ muon. Each group is then reconstructed with a maximum likelihood method that takes into account the scattering of light in ice. After this initial reconstruction, the groups are resplit according to their time residual relative to the muon bundle reconstruction and re-fit. The first splitting forces the activated DOMs into two groups, which is not correct for the background events which don't have a high $p_{T}$ muon and generate only a single shower in the array. Splitting the activated DOMs a second time according to their time residual allows for the possibility for all the activated DOMs to end up in a single group.  Figure \ref{fig_reco} shows the performance of this clustering algorithm. The zenith angle resolution for groups determined using the true simulation information (black, solid lines) is compared to the resolution for groups determined using the clustering algorithm (red, dashed lines) for the muon bundle (top) and high $p_{T}$ muon (bottom). Roughly 20\% of the events fail to reconstruct because there are not enough DOMs in one of the groups.

These reconstruction algorithms achieve a zenith angle resolution of 2.7$^\circ$ for the muon bundle and 4.1$^\circ$ for a high $p_{T}$ muon separated by 400~m. The resolution is worse for the high $p_{T}$ muon because fewer DOMs are activated. While high $p_{T}$ muons with a greater separation are much easier to resolve with the two track algorithm, they also tend to be lower energy (see Eq. \ref{pt_eqn}) and activate fewer DOMs. The average number of DOMs activated by the high $p_{T}$ muon is 12, compared to 50 for the muon bundle. Additionally, because high $p_{T}$ muons have less activated DOMs, a DOM activated by the muon bundle that is incorrectly placed in the high $p_{T}$ group has a much larger effect on the reconstruction of the high $p_{T}$ track (it can also degrade the bundle resolution, but to a lesser extent). These factors lead to a poorer resolution in the high $p_{T}$ track direction.

\section{Signal and Background Separation}
The interaction models in the air shower simulation program CORSIKA \cite{heck} do not include many of the processes that can generate a high $p_{T}$ muon, with the exception of the modified DPMJET model discussed in \cite{berghaus}. Also, since high $p_{T}$ muons will occur in only a fraction of simulated showers making simulation very time intensive, a toy model based on CORSIKA proton showers was used to model the signal. A single muon is inserted into an existing CORSIKA event containing a muon bundle from an air shower. This modified shower is then run through the standard IceCube propagation, detector simulation, and reconstruction routines. Simulations insert a muon with energy of 1 TeV separated 100, 150, 200, and 400~m from the shower core. This gave a clean sample ideal for the development of reconstruction routines optimized to identify simultaneous parallel tracks inside the detector.

\begin{figure}[h]
  \centering
  \includegraphics[width=3.0in]{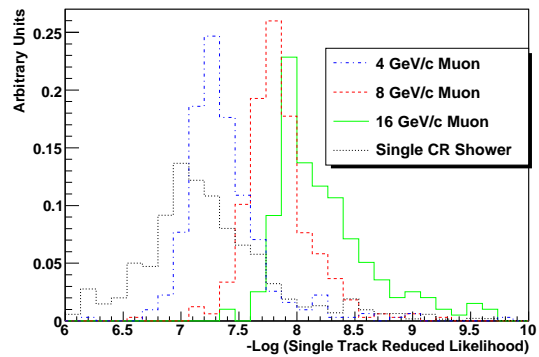}
  \caption{Negative log of the reduced likelihood for the single track reconstruction}
  \label{fig_like}
 \end{figure}

Cosmic rays air showers that do not generate a high $p_{T}$ muon are considered a background to this search. Since they generate only a single shower in the array, these events are mostly eliminated by requiring there be two well-reconstructed tracks in the IceCube detector. These single showers are well-reconstructed by a single track hypothesis, while the high $p_{T}$ muon events are not. Figure \ref{fig_like} shows the negative log of the reduced likelihood of a single track reconstruction for single showers, and showers with an inserted 4, 8, and 16 GeV/c $p_{T}$, 1 TeV muon (separation of 100~m, 200~m, and 400~m from the shower core, assuming an average interaction height of 25~km). Well-reconstructed events have a lower value on this plot. For large separations, this variable separates single showers from showers which contain a high $p_{T}$ muon. When the separation between the high $p_{T}$ muon and the shower core drops below the interstring distance (the blue, dot-dashed line in Figure \ref{fig_like}), it is no longer possible to cleanly resolve the high $p_{T}$ muon from the shower core and the event looks very similar to a single shower.

The IceCube 22-string configuration is large enough that the rate of simultaneous events from cosmic rays is significant. Muon bundles from two (or more) air showers can strike the array within the 10 ${\mu}s$ event window, producing two separated tracks. These so-called double-coincident events are the dominant background for air showers with high $p_{T}$ muons. Since these double-coincident events are uncorrelated in direction and time, requiring that both reconstructed tracks be parallel and occur within 1 ${\mu}s$ can eliminate most of these events. Due to misreconstructions some events will survive this selection, and criteria to eliminate these misreconstructed events are being developed. However, an irreducible background remains from double-coincident events that happen to come from roughly the same direction and time. A conservative estimate of the number of these events is comparable to the number of expected high $p_{T}$ events. A full simulation is under way to determine the rate of double-coincident cosmic ray events and to develop methods to distinguish high $p_{T}$ events from this background.

 \begin{figure}[!h]
  \centering
  \includegraphics[width=1.65in]{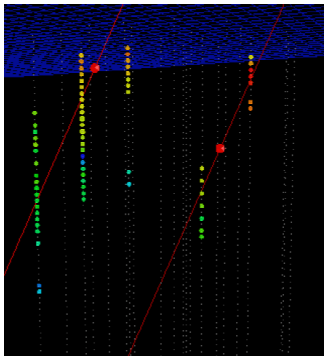}
  \caption{Candidate shower with a high $p_{T}$ muon. The colored circles show DOMs that are activated by light. The color and size of the circle corresponds to the time (red being earliest) and magnitude of the signal, respectively and the white dots indicate DOMs that are not activated in the event. The red lines show the reconstructed tracks (with red circles denoting the arbitrary vertex of the reconstruction).}
  \label{fig_mine}
 \end{figure}
After applying selection criteria to reduce the cosmic ray background, a 10\% sample of the IceCube data was scanned by eye for high $p_{T}$ muon candidates. Several events were found and Figure \ref{fig_mine} shows a representative event. The two tracks occur within 400 ns of each other and the space angle between the two reconstructions is 5.6$^\circ$.

\section{Horizontal Events}
One possible method to avoid the backgrounds from single and double-coincident cosmic ray air showers is to search for events that come from the horizontal direction. At a zenith angle of 60$^{\circ}$ the slant depth through the ice is about 3~km. Only muons with energies in excess of several TeV will reach IceCube \cite{berghaus2} and the event multiplicity is greatly reduced. Because the intra-DOM spacing is 17~m, compared to the intra-string spacing of 125~m relevant for vertical events, the two-track resolution should improve. A disadvantage to searching in the horizontal direction is the increase in the deflection from the magnetic field and multiple scattering with the slant depth, which could lead to a greater spread in the muon bundle and create event topologies which mimic a high $p_{T}$ muon event. Nevertheless, the advantage of background suppression is strong. In addition to high $p_{T}$ muons in cosmic rays, there are a number of other processes (such as the decay of the supersymmetric stau \cite{olivas}) that can produce horizontal parallel tracks, making the horizon an interesting direction.

\begin{figure}[!t]
  \centering
  \includegraphics[width=1.65in]{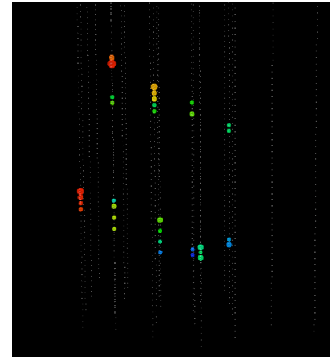}
  \caption{Candidate double track near-horizontal event}
  \label{fig_sandy}
 \end{figure}

A search was conducted in 10\% of the IceCube data from 2007 for these horizontal events using simple topological selection criteria. Events which reconstructed within 30$^\circ$ of horizontal were selected and searched for a double track topology. Several interesting candidate events were found, one is shown in Figure \ref{fig_sandy}.
\section{Conclusions}
IceCube is large enough that the study of high $p_{T}$ muons has become viable. Resolution of tracks separated by at least 150~meters from the shower core will allow identification of muons with $p_{T}$s of at least 6 GeV/c. A two-track reconstruction algorithm has been developed and selection criteria for identification of air showers with high $p_{T}$ muons is under development. A search for horizontal double tracks is also under way. The rate of high $p_{T}$ muon production is sensitive to the composition of the cosmic rays and offers an alternative to existing composition studies.

This work is made possible by support from the NSF and the DOE.

    \end{document}